\documentclass[aps,twocolumn,prb,showpacs,preprintnumbers,amsmath,amssymb]{revtex4}%Physical Review B

\usepackage{amsmath,amssymb}
\usepackage{ascmac}
\usepackage{subfigure}
\usepackage{wrapfig}
\usepackage{pxfonts}
\usepackage{graphicx}% Include figure files
\usepackage{dcolumn}% Align table columns on decimal point
\usepackage{bm}% bold math
\usepackage{txfonts}
\usepackage{color}

\newcommand{\sbra}[1]{\left(#1\right)}
\newcommand{\mbra}[1]{\left\{#1\right\}}
\newcommand{\bbra}[1]{\left[#1\right]}
\newcommand{\tbra}[1]{\left\langle#1\right\rangle}

\newcommand{\iniL}{\left<i\right|}
\newcommand{\iniR}{\left|i\right>}
\newcommand{\vacL}{\left<0\right|}
\newcommand{\vacR}{\left|0\right>}

\newcommand{\ki}{{k_i}}
\newcommand{\kz}{{k_0}}

\newcommand{\tp}{{t^\prime}}

\begin{document}
\title{Transient Carrier Dynamics in a Mott Insulator with Antiferromagnetic Order}

\author{Eiki Iyoda}$^\ast$
\email{iyoda@noneq.c.u-tokyo.ac.jp}
%\email{iyoda@cmpt.phys.tohoku.ac.jp}
\author{Sumio Ishihara}
\affiliation{%
Department of Physics, Tohoku University, 
Sendai 980-8578 Japan}%
\affiliation{%
JST-CREST, Sendai 980-8578 Japan}%
\date{\today}

\begin{abstract}
We study transient dynamics of hole carriers injected at a certain time into a Mott insulator with antiferromagnetic long range order. 
This is termed ``dynamical hole doping" as contrast with chemical hole doping. 
Theoretical framework for the transient carrier dynamics are presented based on the two dimensional $t-J$ model. 
Time dependences of the optical conductivity spectra as well as the one-particle excitation spectra are calculated based on the Keldysh Green's function formalism at zero temperature combined with the self-consistent Born approximation. 
At early stage after dynamical hole doping, the Drude component appears, 
and then incoherent components originating from hole-magnon scatterings start to grow. 
Fast oscillatory behavior due to coherent magnon, and slow relaxation dynamics are confirmed in the spectra. 
Time profiles are interpreted as that doped bare holes are dressed by magnon clouds, and are relaxed into spin polaron quasi-particle states. 
Characteristic relaxation times for Drude and incoherent peaks strongly depend on momentum of a dynamically doped hole, and the exchange constant. 
Implications to the recent pump-probe experiments are discussed. 
\end{abstract}

\maketitle

%%%%%%%%%%%%%%%%%%%%%%%%%%%%%%%%
%%%%%%%%%%%%%%%%%%%%%%%%%%%%%%%%
\section{Introduction}

Ultrafast carrier dynamics in strongly correlated electron systems have significantly attracted much attention, because a number of time-resolved experimental technique are rapidly developed in the last decade. 
In contrast to conventional metals and semiconductors, correlated electron systems show a rich variety of competing phases, interactions and degrees of freedom. 
Intensive external stimuli, such as short laser pulse, DC and AC electric fields, trigger a breaking of a subtle balance between them, and often induce transiently some non-trivial states, which are never realized in thermal equilibrium states. 
A number of experiments and theoretical analyses for transient carrier dynamics 
have been done in several classes of correlated electron systems, e.g. charge-ordered organic salts, 
spin cross-over complexes, magnetic oxides, multiferroics and superconducting materials.~\cite{Nasu,JPSJ}

It is widely recognized that one of the prototypical and not yet revealed issues in transient electron dynamics is photo-irradiation effects in a Mott insulator with an antiferromagnetic long-range order (AFLRO). 
This is motivated not only from deep understandings of a Mott insulator and the high-Tc superconductivity (HTSC), but also from a search for drastic photoinduced non-equillibrium states.~\cite{Oka,Heidrich,Vidmar1,Mierzejewskii,Freericks, Morits, Werner}
The optical pump-and-probe experiment is a standard method which directly detects transient electron and hole dynamics excited across a Mott gap.~\cite{Han,Matsuda,Cavalleri,Iwai1,Kubler,Perfetti,Takahashi,Tsuji,Filippis,Matsueda,Kanamori1,Kanamori2,Ohara}
Ultrashort optical pulses have revealed real time dynamics of electrons and holes, and their relaxations toward recombination and thermalization. 
Recent pump-probe experiments in quasi-two dimensional Mott insulators, as parent compounds of HTSC, revealed that a metallic Drude component appears just after photoexcitation and decays within several tens femtosecond.~\cite{Okamoto1} 
After several picoseconds, the midgap absorptions appear, and two absorption peaks are separately identified as hole and electron carrier contributions. 
These systematic experimental observations require detailed understanding of carrier dynamics doped temporally into a Mott insulator with AFLRO. 

\begin{figure}
\begin{center}
\includegraphics[width=6.0cm]{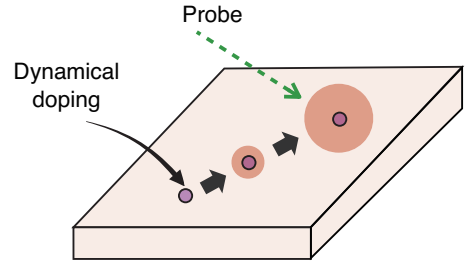}
\caption{A schematic dynamical hole doping.}
\label{fig:dynamical_doping}
\end{center}
\end{figure}
\par
In this paper, we examine real-time dynamics of hole carriers introduced at a certain time into a Mott insulating state with AFLRO. This doping is termed ``dynamical hole doping", and the introduced holes are ``dynamically doped holes", in this paper, as contrast with chemically doped holes (see Fig.~\ref{fig:dynamical_doping}).
The dynamical hole doping, not electron-hole pairs, is not only for a simple theoretical setup, but also have implications to the recent optical pump-probe experiments, where dynamics of photo-doped holes and electrons were able to be separated.~\cite{Okamoto1,Hiroi,Muraoka,Yada1,Yada2} 
We present a theoretical framework for the dynamically doped holes base on the two-dimensional $t-J$ model. 
We adopt the Keldysh Green's function formalism~\cite{Rammer,Kamanev} and the self-consistent Born (SCB) approximation,~\cite{Kane,Martinez}  which is known to describe well motions of chemically doped holes in a Mott insulator with AFLRO.
To examine real time dynamics of the dynamically doped holes, the transient optical conductivity spectra, as well as the one-particle excitation spectra, are calculated. 
At early stage after hole doping, the Drude component is only appears in the optical conductivity spectra, 
and then, incoherent peaks originating from hole-magnon scatterings start to grow. 
Intensities in these peaks at early stage are proportional $t^2$ and $t^3$. 
Fast oscillatory behavior due to coherent magnon, and slow relaxation dynamics are confirmed in the spectra. 
Time profiles are interpreted as that doped bare holes are dressed by magnon clouds, and are relaxed into spin polaron quasi-particle in steady state. 
We show that relaxation time scales strongly depend on momentum of the dynamically doped hole, and magnitude and anisotropy of the exchange constant. 

In Sec.~II, the $t-J$ model represented by the spinless fermion and spin wave operators are introduced. 
We present, in Sect. ~II, theoretical formalism for real-time dynamics of doped holes based on the Keldysh Green's function and the SCB approximation. 
Numerical results of transient hole dynamics are given in Sec.~IV. 
Section V is devoted to conclusion. 
In Appendix A, we present the dominant pole approximation, adopted to analyze the numerical results in Sec.~IV.  

%%%%%%%%%%%%%%%%%%%%%%%%%%%%%%%%
%%%%%%%%%%%%%%%%%%%%%%%%%%%%%%%%
\section{Model}

We start from the $t-J$ model in a two-dimensional square lattice given by 
\begin{align}
{\cal H}&=-t_0\sum_{\langle ij \rangle  \sigma} 
\left ( {\widetilde c}_{i \sigma}^\dagger {\widetilde c}_{j \sigma} +H.c. \right )  \nonumber \\
&+\sum_{\langle ij \rangle} 
\left \{ J_{\parallel} S_i^z  S_j^z +\frac{J_{\perp}}{2} \left ( S_i^+  S_j^- + S_i^- S_j^+\right ) \right \} , 
\label{eq:tjH} 
\end{align}
where ${\widetilde c}_{i \sigma}=c_{i \sigma}(1-n_{i -\sigma})$ is an annihilation operator for an electron with spin $\sigma(=\uparrow, \downarrow)$ at site $i$ without double occupancy, 
and ${\bf S}_i$ is a spin operator with amplitude of $S=1/2$. 
Transfer integral is represented by $t_0$ to distinguish from a simbol of time.
For convenience, we introduce an anisotropy in the exchange interaction as a parameter $\alpha \equiv J_{\perp}/J_{\parallel}$; $\alpha=1$ ($\alpha=0$) corresponds to the Heisenberg (Ising) limit. 

We assume a AFLRO state in the ground state, where sublattices for up- and down-spins are termed A and B, respectively. 
Spin operators are represented by magnon operators introduced by the Holstein-Primakov transformation as 
$S_i^+=(1-a_i^\dagger a_i)^{1/2} a_i$,  
$S_i^-=a_i^\dagger (1-a_i^\dagger a_i)^{1/2}$ and $S_i^z=1/2-a_i^\dagger a_i$ for sublattice A 
and similar ways for sublattice B. 
By following Refs.~\onlinecite{Kane, Martinez}, the electron operator without double occupancy is given by the slave fermion representation as 
$\widetilde c_{i \sigma}^\dagger=h_i a_{i \sigma}^\dagger$ 
with the constraint 
$\sum_\sigma a_{i \sigma}^\dagger a_{i \sigma}+h_i^\dagger h_i=1$ 
where $h_i$ is a spinless fermion operator for hole. 
Up to the lowest order of the $1/S$ expansion, 
where $a_{i\in A \downarrow}$ and  $a_{i\in B \uparrow}$ are replaced by $\sqrt{2S}(=1)$, 
the hopping term of the Hamiltonian is given as 
\begin{align}
{\cal H}_t=-t_0 \sum_{\langle ij \rangle i\in A j\in B} h_i h_j^\dagger \left( a_{i \uparrow}^\dagger+a_{j \downarrow} \right) +H.c.  . 
\label{eq:Ht0}
\end{align}
The $J$-term expressed by the bilinear form of the boson operator is diagonalized by the Bogoliubov transformation. 

By introducing the Fourier transformation, 
the Hamiltonian is finally given by 
\begin{align}
{\cal H}_t
=\frac{zt_0}{\sqrt{N}}\sum_{\bm{k,q}}
h_{\bm k}^\dag h_{\bm{k-q}}\alpha_{\bm q} M_{kq}+H.c. , 
\label{eq:Ht}
\end{align}
for the $t$-term with a coupling constant 
\begin{align}
M_{kq}=u_q\gamma_{k-q}+v_q\gamma_k,
\label{eq:mkq}
\end{align}
and 
\begin{align}
{\cal H}_J=\sum_{\bm q}
\omega_{\bm q} \alpha^\dag_{\bm q} \alpha_{\bm q} , 
\label{eq:Hj}
\end{align}
for the $J$-term. 
Here, $\alpha_{\bm q}$ is a boson operator for magnon introduced by the Bogoliubov transformation, and $h_{\bm q}$ is the Fourier transform of $h_i$. 
We define the magnon dispersion 
$\omega_q=zJS(1-\delta_d)^2\nu_q$
where $\delta_d$ is a hole density, $S=1/2$, 
and $\nu_q=[1-(\alpha\gamma_q)^2]^{1/2}$
with the form factor 
$ \gamma_q=(\cos q_x+\cos q_y)/2$, 
and the coordination number $z(=4)$. 
Factors $u_q$ and $v_q$ in $M_{kq}$ are given by the usual expression in the Bogoliubov transformation defined by 
$u_q=[(1+\nu_q)/(2\nu_q)]^{1/2}$
$v_q=-sgn(\gamma_q) [(1-\nu_q)/(2\nu_q)]^{1/2}$. 
As well known, there is the absence of a free kinetic energy for the spin-less fermion, 
and a coupling $M_{kq}$ between a fermion and magnons in the $t$-term is the source of hole dynamics. 
From now on, we adopt units of energy and time as $t_0$ and $t_0^{-1}$, respectively.

%%%%%%%%%%%%%%%%%%%%%%%%%%%%%%%%
\section{Formulation}
\label{sec:formulation}

%%%%%%%%%%%%%%%%%%%%%%%%%%%%%%%%
\subsection{Initial state and Keldysh formalism}
\label{sec:initial}

We explain a situation of the dynamical hole doping and how to observe transient states. 
First, a hole is introduced in the N$\rm \acute{e}$el ordered state at half filling by an external field at $t=0$.  
This state is adopted as an initial state, and is simulated by 
\begin{align}
\left|i\right>=\sum_{\ki} g(\ki )h_{\ki}^\dag\left|0\right>,
\label{eq:initial_state}
\end{align}
where $|0\rangle$ expresses the N$\rm \acute{e}$el state, and is termed ``vacuum". 
A function $g(\ki)$ describes a momentum distribution of a doped hole 
and its functional form depends on actual hole-injection way. 

\begin{figure}
\begin{center}
\includegraphics[width=6.0cm]{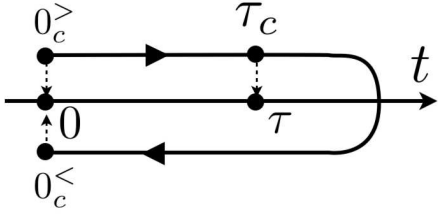}
\caption{A Keldysh contour.}
\label{fig:Kcontour}
\end{center}
\end{figure}

This initial state is time-evolved by the Hamiltonian ${\cal H}_t+{\cal H}_{J}$ given in Eqs.~(\ref{eq:Ht}) and (\ref{eq:Hj}). At time $t>0$, the expectation value of an operator is defined as 
\begin{align}
\tbra{O(t)}=\iniL O_H(t)\iniR,
\end{align}
where $O_H(t)$ is an operator $ O$ in the  Heisenberg representation.
In order to use the Wick's theorem, we introduce the Keldysh's closed-time contour whose length is finite, 
as shown in Fig.~\ref{fig:Kcontour}. 
Then, $\tbra{O(t)}$ is given by 
\begin{align}
\tbra{O(t)}
=
\sum_{\ki} g(\ki)^2
\vacL T_C
\left\{
U_C
h_{\ki}(0_c^<)
O(t_c^<)
h_{\ki}^\dag(0_c^>)
\right\}
\vacR ,
\end{align}
with 
\begin{align}
U_C
=
T_C
\exp\sbra{\int_C -i {\cal H}_t(t_c^{\prime})dt_c^{\prime}}, 
\end{align}
where all time-dependent operators are represented in the interaction representation.
Time variable on a Keldysh's contour is denoted by $t_c$ with a subscript $c$, and 
a projection of $t_c$ onto the real time is $t$. 
An upperscript $>$ ($<$) represents that the contour time is on the upper (lower) branch of the Keldysh's contour.
A symbol $T_C$ is a Keldysh's contour ordering operator, 
and $\int_C$ implies an integral along the contour. 
We have relations $0_c^< >_K t_c$ and $0^{>}_c <_K t_c$ for any $t_c$,
where $<_K$ and $>_K$ represent inequalities defined on the Keldysh's contour.
The Wick's theorem and the Feynman's rules are applicable to this form of the expectation value. 
We note that the imaginary contour from $t=0$ to $-i/(k_{\rm B}T)$, 
known in the  Kadanoff-Baym's L-shaped contour, is unnecessary to be taken into account 
in the present formalism. 
This is because the initial state is explicitly given in Eq.~(\ref{eq:initial_state}), 
and thermal equilibrium processes along the imaginary contour is not necessary. 

%%%%%%%%%%%%%%%%%%%%%%%%%%%%%%%%
\subsection{Green's function}
\label{sec:gf}

In the Keldysh formalism, one-particle Green's function (non-equilibrium Green's function)  for holes is defined as
\begin{align}
G(k; t_c,t_c^\prime)
&=
-i\tbra{
T_C
\mbra{
h_k(t_c)h_k^\dag(t_c^\prime)}
},
\label{eq:GF}
\end{align}
which is rewritten for the initial state introduced above as 
\begin{align}
G(k; t_c,t_c^\prime)
&=
-i \sum_{\ki} g({\ki})^2
\vacL T_C
\left\{
U_C
%\exp\sbra{\int_C -i H_t(t_c^{\prime\prime})dt_c^{\prime\prime}}
h_{\ki}(0_c^{<})
h_k(t_c)
h_k^\dag(t_c^\prime)
h_{\ki}^\dag(0_c^{>})
\right\}
\vacR . 
\end{align}
The real-time Green's functions are introduced by 
\begin{align}
G_{ij}(k; t,\tp)=G(k; t^{a_i},\tp^{a_j}),
\end{align}
where $a_1$ and $a_2$ take $>$ and $<$.
Here, $G_{11}$ and $G_{22}$ are the causal and anti-causal Green's functions, respectively, 
and $G_{12}$ and $G_{21}$ are the lesser and greater Green's functions, characterizing the particle distribution, respectively. 
From these components, 
the retarded Green's function is given by 
\begin{align}
G^R(k; t,t^\prime)&=G_{11}(k; t,\tp)-G_{12}(k; t,\tp) \nonumber \\
&=G_{21}(k; t,\tp)-G_{22}(k; t,\tp) .
\end{align}
One-particle excitation spectra for holes are defined by
\begin{align}
A(k; \omega,t^\prime)=-\frac{1}{\pi}\mathrm{Im}\bbra{G^R(k; \omega,\tp)} , 
\end{align}
where we define the Fourier transform of the real-time Green's function given by 
\begin{align}
G(\omega,t^\prime)=\int d(t-t')e^{i\omega (t-t')}G(t-t^\prime,t^\prime).
\end{align}

In addition, we introduce one-particle Green's functions defined as expectations with respect to the vacuum state $\vacR$ as
\begin{align}
g(k; t_c, t_c^\prime)
=
-i\vacL T_C\mbra{
U_C
h_k(t_c)h_k^\dag(t_c^\prime)
}\vacR,
\end{align}
for a hole, and 
\begin{align}
d(q; t_c, t_c^\prime)
=
-i\vacL T_C\mbra{
U_C
\alpha_q(t_c)\alpha_q^\dag(t_c^\prime)
}\vacR,
\end{align}
for a magnon.  
The real-time Green's functions, $g(k;t-t^\prime)$ and $d(q;t-t^\prime)$, are obtained by projections on to the real-time axis, and depend only on the time difference $t-\tp$.
The lesser component $g_{12}(k;t,\tp)=0$, because of no holes in the vacuum state.
We will show later that $G(t_c,t_c^\prime)$ in Eq.~(\ref{eq:GF}) is given by combinations of 
$g(k;t_c, t_c^\prime)$ and $d(k;t_c, t_c^\prime)$, and $g(k;t_c, t_c^\prime)$ is calculated by the self-consistent Born approximation explained in Sect.~\ref{sec:scb}.

\begin{figure}
\begin{center}
\includegraphics[width=9.0cm]{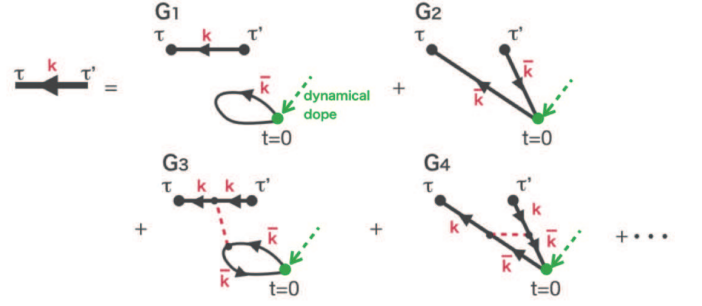}
\caption{Feynman diagrams for the one-particle Green's functions. 
Solid and broken lines represent the hole Green's functions, $g(k; t)$, and the magnon Green's function, $d(k; t)$, respectively, defined in the vacuum state.
}
\label{fig:Gexpand}
\end{center}
\end{figure}
The Green's function $G(t_c,t_c^\prime)$ in Eq.~(\ref{eq:GF}) is a four-point function, 
and is evaluated by the perturbational expansion with respect to $M_{kq}$.
Figure~\ref{fig:Gexpand} shows the Feynman diagrams in a series expansion up to the second order of $M_{kq}$.
The zeroth order terms, $G_1$ and $G_2$, are represented by products of 
$g(t_c,t_c^\prime)$ as 
$G_1 \sim -ig(t_c,t_c^\prime)g(0_c^<,0_c^>)$ and $G_2 \sim ig(t_c,0_c^>)g(0_c^<,t_c^\prime)$. 
In the second order terms, $G_3$ and $G_4$, 
the two Green's functions are connected by the magnon Green's function. 
Diagrams for $G_1$ and $G_3$ represent ``direct" processes, and 
$G_2$ and $G_4$ represent ``exchange" processes where a dynamically doped hole is exchanged with an additionally introduced hole. 

Explicit forms of the retarded components are given as 
\begin{align}
G_1^R(k; t,t^\prime)
=
g^R(k;t-t^\prime),
\label{eq:g1}
\end{align}
\begin{align}
G_2^R(k; t,t^\prime)
=
0,
\label{eq:g2}
\end{align}
\begin{align}
G_3^R(k; t,t^\prime)
&=
\sum_{\ki} g({\ki})^2
\frac{M_{k0}M_{\ki 0}}{2}
\int_\tp^t dt_2
\int_0^{t_2}
dt_3 \nonumber \\
&\times 
g^R(k;t-t_2)g^R(k;t_2-t^\prime) 
\left|g^R(\ki;t_3)\right|^2 \nonumber \\
& \times
\sbra {d^R(0;t_2-t_3)+d^{R*}(0;t_2-t_3) },
\label{eq:g3}
\end{align}
\begin{align}
G_4^R(k; t,t^\prime)
&=\sum_{\ki} g({\ki})^2
\frac{-1}{2}
\int_\tp^t dt_2
\int_\tp^{t_2}
dt_3 g^R(k; t-t_2)g^R(\ki;t_2) \nonumber \\
&\times
g^{R*}(\ki;t_3)g^R(k; t_3-\tp) \nonumber \\
&\times
\sbra{M_{k,k-\ki}^2 d^R(k-\ki;t_2-t_3)+
M_{\ki,\ki-k}^2 d^{R*}(\ki-k;t_2-t_3)}.
\label{eq:g4}
\end{align}
As shown in Eq.~(\ref{eq:g1}), $G_1$ does not depend on the initial state and is nothing but an one-particle Green's function in a Mott insulator, 
since additionally doped hole does not interact with an dynamically doped hole. 
We have $G_2^R=0$, because $G_2$ includes the so-called ``an electron line", $g(0,\tp>0)$, 
which vanishes in the $t-J$ model at half-filling. 
The remaining terms, $G_3$ and $G_4$, explicitly depend on double times, $t$ and $t^\prime$, 
and characterize the transient processes. 

%%%%%%%%%%%%%%%%%%%%%%%%%%%%%%%%
\subsection{Optical conductivity}
\label{sec:oc}

To calculate the optical conductivity in a transient state, 
we set up the Hamiltonian ${\cal H}({\bm A})$ where 
the electric field is introduced as a Peierls phase in the $t$-term 
of the Hamiltonian defined in Eq.~(\ref{eq:Ht}). 
The current operator is given as 
\begin{align}
{\bm j}=-c\frac{\partial {\cal H}({\bm A})}{\partial {\bm A}} 
=-\frac{ie t_0}{\hbar}\sum_{<ij>} {\bm \delta}_{ij}
\left (a_j^\dag h_i^\dag h_j-h_j^\dag h_i a_j \right)  ,
\end{align}
where ${\bm A}$ is a vector potential, ${\bm \delta}_{ij}$ 
is a vector connecting NN sites $i$ and $j$. 
Here, we take $e=1$, $c=1$ and $\hbar=1$, for simplicity. 
By applying the Bogoliubov and Fourier transformations, the current operator is rewritten as 
\begin{align}
{\bm j}=-\frac{1}{\sqrt{N}}\sum_{kq} {\bm V}_{kq} j_{kq},
\label{eq:jv}
\end{align}
where we define 
\begin{align}
j_{kq}=\alpha^\dag_q h_{k-q}^\dag h_k+h_k^\dag h_{k-q}\alpha_q,
\label{eq:jkq}
\end{align}
and 
\begin{align}
{\bm V}_{kq}=u_q {\bm v}_{k-q}+v_q{\bm v}_k,
\label{eq:vkq}
\end{align}
with 
${\bm v}_k=\partial \epsilon_k/\partial k$ and 
$\epsilon_k=-(zt_0/2)(\cos a k_x+\cos a k_y)$. 

The optical conductivity is formulated within a linear response regime for 
a electric field ${\bm E}(t)={\bm E} e^{-i\nu t}$ for probe photons where $\nu$ is a photon frequency. 
Optical conductivity is defined as a response function for an electric current $\mathcal{{\bm J}}(t) \equiv \tbra{{\bm j}(t)}$ at time $t$ induced by an electric field ${\bm E}(t')$  at time $t'$. 
This is given as 
\begin{align}
\left.\frac{\partial \mathcal{J}_\alpha(t)}{\partial E_\beta}\right|_{{E}=0}
=
\int_{-\infty}^t dt^\prime \sigma_{\alpha\beta}^{(\nu)}(t,t^\prime)e^{-i\nu t^\prime} , 
\end{align}
where $E_{\beta}=[{\bm E}]_\beta$.
From Eq.~(\ref{eq:jv}), the left hand side is divided into the two parts as 
\begin{align}
\frac{\partial\mathcal{J}_\alpha}{\partial E_\beta}
=
-\frac{1}{\sqrt{N}}\sum_{kq}
\left ( 
\frac{\partial {V}^\alpha_{kq}}{\partial E_\beta}
\tbra{j_{kq}(t)}
+
{ V}^\alpha_{kq}
\frac{\partial}{\partial E_\beta}
\tbra{j_{kq}(t)}
\right )  .
\end{align}
where $ {V}^\alpha_{kq}=[ {\bm V}_{kq}]_\alpha$.
This is calculated from Eqs.~(\ref{eq:jkq}) and (\ref{eq:vkq}) ,  
and then the optical conductivity is explicitly given as 
\begin{align}
\sigma_{\alpha\beta}^{(\nu)}(t,t^\prime)
&=-\frac{1}{\nu}
\mbra{
\delta(t-t^\prime)i\tbra{ \mathcal{E}^{\alpha \beta}(t)}
+
\theta(t-t')
\tbra{
\left[
j_\alpha(t),j_\beta(t^\prime)
\right]
}
},
\label{eq:nelkubo}
\end{align}
where $\mathcal{E}^{\alpha \beta}$ is the energy stress tensor defined by 
\begin{align}
\mathcal{E}^{\alpha\beta}
&=
\frac{zt}{\sqrt{N}}\sum_{kq} {\widetilde M}_{kq}^{\alpha\beta}
(h_k^\dag h_{k-q}\alpha_q+\alpha^\dag_q h_{k-q}^\dag h_k),
\end{align}
and 
\begin{align}
{\widetilde M}_{kq}^{\alpha\beta}
&=
\delta_{\alpha\beta} \sbra{u_q {\widetilde \gamma}^\alpha_{k-q}+v_q {\widetilde \gamma}_k^\alpha} ,
\end{align}
with ${\widetilde \gamma}_k^\alpha=\sbra{\cos k_\alpha}/2$.
The first and second terms in Eq.~(\ref{eq:nelkubo}) represent the diamagnetic and paramagnetic components, respectively.
We assume no electric current before applying the electric field. 
The paramagnetic component is given in the Keldysh formalism by 
\begin{align}
\tbra{\bbra{j_\alpha(t),j_\beta(t^\prime)}}
=
\sum_{a=<,>}(\delta_{a,<}-\delta_{a,>})
\tbra{
j_\alpha(t_c^<)j_\beta(t_c^{\prime a})
} . 
\label{eq:current_com}
\end{align}
For convenience, we introduce the current-current response function as 
\begin{align}
\chi_{\alpha\beta}(t,t^\prime)
=
-i\theta(t-\tp) 
\tbra{\bbra{j_\alpha(t),j_\beta(t^\prime)}} .
\end{align}

We define the optical conductivity spectra at time $t$ by introducing the Fourier transformation as 
\begin{align}
\sigma_{\alpha \beta}(\omega,t)
=
\int_{-\infty}^{\infty} d(t-t')e^{i\omega (t-\tp)}\sigma_{\alpha \beta}^{(\nu)}(t,\tp) . 
\end{align}
This has a physical meaning only at $\omega=\nu$, 
because 
\begin{align}
\mathcal{J}_\alpha(t)
&=
\int_{-\infty}^t
d\tp 
\sigma_{\alpha\beta}^{(\nu)}
(t,t^\prime)
E_\beta e^{-i \nu \tp } \nonumber
\\
&=
\sigma_{\alpha\beta}^{(\nu)}
(\omega=\nu,t)E_\beta(t),
\end{align}
where a relation $\sigma_{\alpha\beta}^{(\nu)}(t, \tp)=0$ for $t-\tp<0$ is used. 
The real part of the optical conductivity is given by 
\begin{align}
\mathrm{Re}\sigma(\omega,t)
&=
\pi\delta(\omega)D_{\alpha\beta}(t)-\frac{\omega}{\omega^2+\eta^2}\mathrm{Im}\chi_{\alpha\beta}(\omega,t),
\label{eq:red}
\end{align}
with an infinitesimal positive constant $\eta$.
The first term is the Drude-component which is divided into the two parts as 
\begin{align}
D_{\alpha\beta}(t)
&=
D^{\mathrm{dia}}_{\alpha\beta}(t)+D^{\mathrm{para}}_{\alpha\beta}(t),
\label{eq:drude}
\end{align}
with the diamagnetic component given by 
\begin{align}
D^{\mathrm{dia}}_{\alpha\beta}(t)
&=
-\tbra{\mathcal{E}^{\alpha\beta}(t)},
\end{align}
and the paramagnetic one by 
\begin{align}
D^{\mathrm{para}}_{\alpha\beta}(t)
&=
\mathrm{Re}\chi_{\alpha\beta}(\omega=0,t) . 
\end{align}

\begin{figure}
\begin{center}
\includegraphics[width=5.0cm]{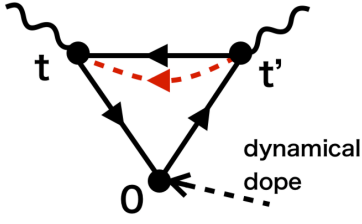}
\caption{A Feynman diagram for the optical conductivity.
Solid and broken lines represent the hole Green's function, $g(k; t)$, and the magnon Green's function, $d({q}; t)$, respectively, defined in the vacuum state, 
and wavy lines are for the light.}
\label{fig:sigma_diagram}
\end{center}
\end{figure}

We calculate the optical conductivity by the perturbational expansions. 
Within the lowest order term, the diamagnetic and paramagnetic components are given by 
\begin{align}
\label{eq:OCdia}
D^{\mathrm{dia}}_{\alpha\beta}(t)=
\frac{2}{N}\sum_{{\ki}, q} g({\ki})^2
\int
dt_2 {\widetilde M}_{\ki, q}^{\alpha\beta}M_{\ki, q}h_{{\ki}, q}(t,t_2) ,
\end{align}
and 
\begin{align}
\chi_{\alpha\beta}(t,t^\prime)
&=
-\frac{2}{N}
\sum_{{\ki}, q} g({\ki})^2
V_{\ki, q}^\alpha V_{\ki, q}^\beta h_q(t,t^\prime) ,
\label{eq:OCpara}
\end{align}
respectively. 
Here we define 
\begin{align}
h_{{\ki}, q}(t,t^\prime)
&=
\mathrm{Im}
\bbra{
g^{R*}(\ki;t)g^R(\ki-q;t-t^\prime)g^R(\ki,\tp)d^R(q;t-\tp)
} . 
\label{eq:sigma_h}
\end{align}
Both contributions are diagramatically expressed by Fig.~\ref{fig:sigma_diagram}.

It is worth to note that ${\rm Re}\sigma(\omega, t)$ for $\omega  \geq t^{-1} $ has a physical meaning of the response function because of the uncertainly principle. 
In the pump-probe experiments in real materials, there are several contributions to damping factors in $\sigma(\omega, t)$, which are not taken into account in the present formalism, such as impurities and vacancies. 
These extrinsic factors change a delta-function for the Drude peak into a Lorentzian peak $\Gamma/(\omega^2+\Gamma^2)$, which is relevant as a response function for a condition of $t \ge 1/\Gamma$.

%%%%%%%%%%%%%%%%%%%%%%%%%%%%%%%%
\subsection{Self-consistent Born approximation}
\label{sec:scb}

As shown in Eqs.~(\ref{eq:g1})-(\ref{eq:g4}) and Eqs.~(\ref{eq:OCdia})-(\ref{eq:sigma_h}), 
the full Green's functions and the optical conductivity are represented by $g^R(k;t)$.
We evaluate $g^R(k;t)$ by using the self-consistent Born approximation, 
which is known to describe well chemically doped hole dynamics in AFLRO, in particular, for $\alpha \sim 1$.~\cite{Kane,Martinez} 

\begin{figure}[t]
\begin{center}
\includegraphics[width=3.0cm]{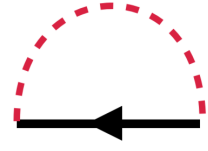}
\caption{
A self energy of a hole Green's function in the self-consistent Born approximation.
Solid and broken lines represent the hole Green's function, $g(k; t)$, and the magnon Green's function, $d(k; t)$, respectively, defined in the vacuum state.
}
\label{fig:self_energy}
\end{center}
\end{figure}

We iteratively solve the Dyson's equation given by 
\begin{align} 
{g^R}^{-1}(k;\omega)
&=
{g^{R(0)}}^{-1}(\omega)-\Sigma^R(k;\omega),
\end{align}
where the self-energy is given by  
\begin{align}
\Sigma^R(k;\omega)
&=
i\frac{z^2t^2}{N}\sum_q\int \frac{d\omega^\prime}{2\pi} M_{kq}^2 g^R(k-q;\omega-\omega^\prime)d^R(q;\omega^\prime),
\label{eq:scba}
\end{align}
which is diagramatically shown in Fig.~\ref{fig:self_energy}.
The bare hole-Green's function is given by 
\begin{align}
{g^{R(0)}}(\omega)
=\frac{1}{
\omega+i\eta }. 
\end{align}
As for the magnon Green's function, 
for simplicity, it is replaced by the bare magnon-Green's function: 
\begin{align}
{d^{R}}(q;\omega)
=\frac{1}{\omega-\omega_q+i\eta} . 
\end{align}
Although the magnon Green's function can be estimated by the series expansion, 
the lowest-order self-energy of the bubble type vanishes, since 
the so-called electron line, where the electron creation operator acts  on a vacuum, 
disappears in the $t-J$ model.  

%%%%%%%%%%%%%%%%%%%%%%%%%%%%%%%%
\section{Numerical Results}
%%%%%%%%%%%%%%%%%%%%%%%%%%%%%%%%
\label{sec:results}

In the numerical calculations, we adopt two dimensional square-lattices with $N \times N$ sites, where $N$ is taken to be 32 in most of the calculations, and $128$ in Fig.~\ref{fig:akw_el}.
A typical number of mesh for energy is 512.
We assume, for simplicity, that the momentum in the dynamically doped hole is a specific value of $k_0$, i.e. $g(k_i)=\delta_{k_i k_0}$ in Eq.~(\ref{eq:initial_state}), and that holes with momentum $k_i$ of density $\delta_d$ are independently injected into a system. 
We take $\delta_d=0.01$, which is realistic as a photo-doped carrier density in optical pump-probe experiments.
All energies and times in the numerical calculations are measured by $t_0$ and $t_0^{-1}$, respectively. 
We adopt the first Brillouin zone for the original square lattice, not the reduced one for the N$\rm \acute e$el state. 

\subsection{Electronic state before dynamical doping}

\begin{figure}[t]
\begin{center}
\includegraphics[width=\linewidth]{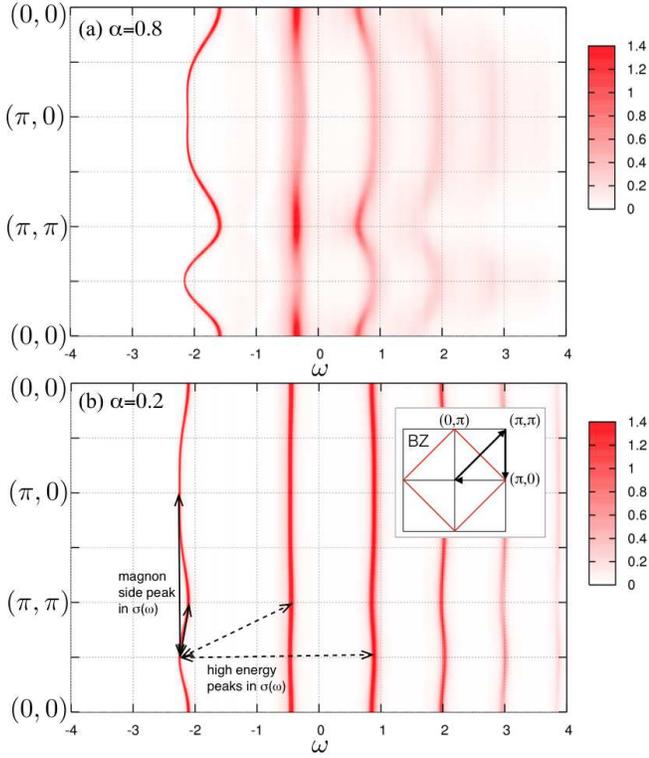}
\caption{
One particle excitation spectra before dynamical doping, $a(k; \omega)=-(1/\pi){\rm Im} g^R(k; \omega)$. 
Parameters are chosen to be $N=128$, 
$J=0.4$, and a damping constant $\eta=0.01$. 
Anisotropy parameter is $\alpha=0.8$ in (a), and $\alpha=0.2$ in (b).
Arrows in (b) represent examples of transitions characterizing peaks in optical conductivity in transient states
(see section \ref{sec:OCinTS}).
}
\label{fig:akw_el}
\end{center}
\end{figure}
\begin{figure}[t]
\begin{center}
\includegraphics[width=0.9\linewidth]{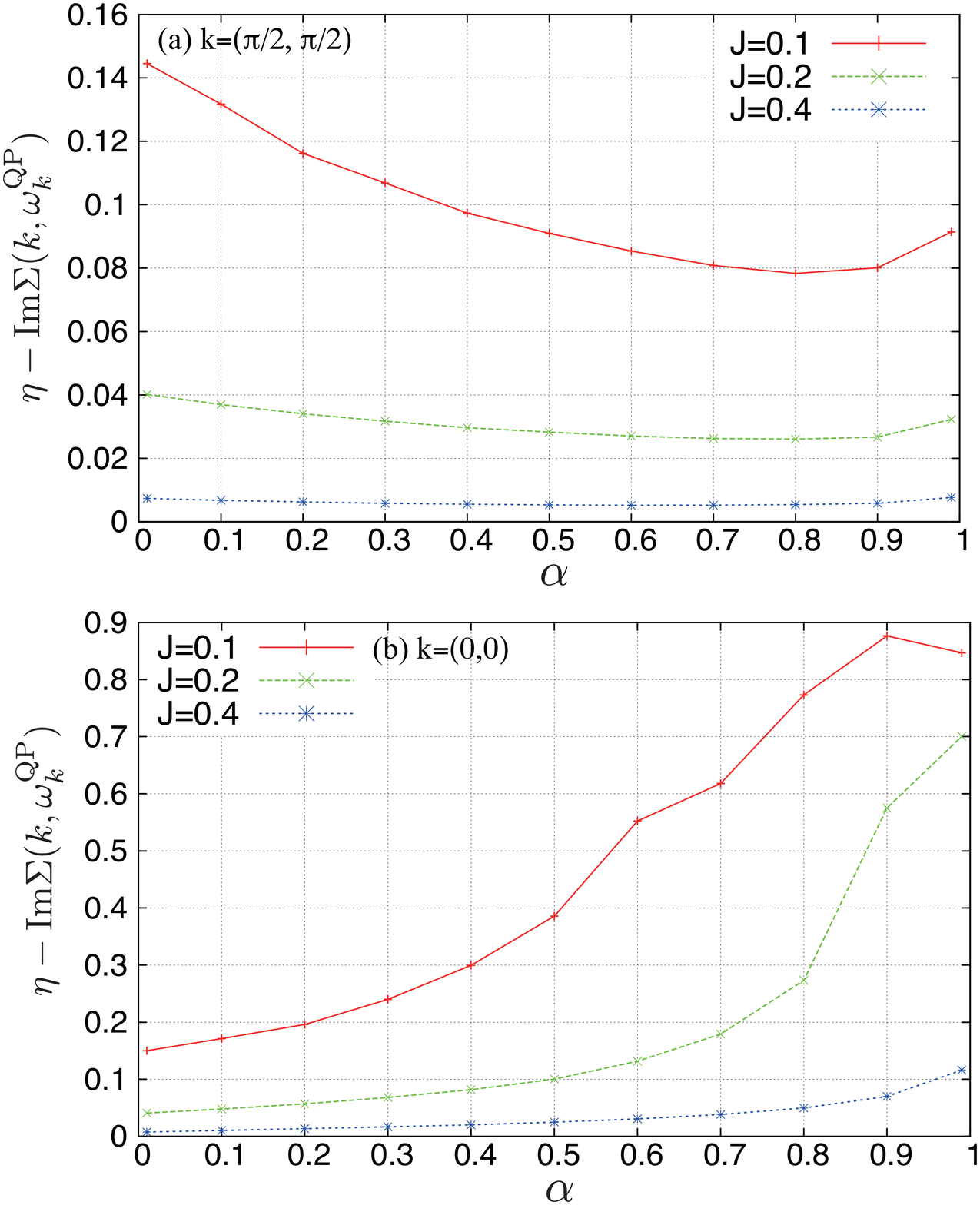}
\caption{Imaginary parts of the self-energy at the QP band for several $J$, 
$- \mathrm{Im}\Sigma(k, \omega^{\rm QP}_{k})+\eta$,  
where $\omega^{\rm QP}_k$ is the QP band energy at momentum $k$. 
Momentum as a parameter in the Green's function is chosen to be $k=(\pi/2, \pi/2)$ in (a), and $k=(0, 0)$ in (b). 
}
\label{fig:sigma_el}
\end{center}
\end{figure}

Before showing the transient electronic state, 
we first show results for one-particle excitation spectra before dynamical doping. 
The present results are consistent with the previous publications in Refs.~\onlinecite{Kane} and \onlinecite{Martinez}. 
Figure \ref{fig:akw_el} shows contour maps of the one-particle excitation spectra given by 
\begin{align}
a(k; \omega)=-\frac{1}{\pi}\mathrm{Im}\  g^R(k; \omega) ,
\end{align}
in the $k-\omega$ planes. 
We chose an anisotropy parameter as $\alpha(\equiv J_\perp/J_\parallel)=0.8$ and $0.2$ in Fig.~\ref{fig:akw_el} (a) and (b), respectively. 
Ladder-like multiple-peak structures are seen in  the Ising-like case [Fig.~\ref{fig:akw_el} (b)] 
where low-lying peaks are separated by $J^{2/3}$, as known in Ref.~\onlinecite{Kane}. 
In a case of large $\alpha$, incoherent background and dispersive 
character due to spin fluctuation are remarkably seen. 
The lowest dispersive branch centered at $\omega \simeq -2$ is the quasi-particle (QP) band, 
where a weak dispersion is seen along the $(\pi, 0)-(\pi/2, \pi/2)$ line, 
and a shallow minimum appears at $(\pi/2, \pi/2)$, i.e. small hole pockets around $(\pi/2, \pi/2)$. 
A large QP dispersion in Fig.~\ref{fig:akw_el}(a) implies that 
carrier propagation in a AFLRO background is owing to spin fluctuation. 

Figure \ref{fig:sigma_el} shows $\alpha$ dependences of the imaginary parts of the self-energy at the QP peak for several values of $J$, i.e. $-\mathrm{Im}\Sigma(k; \omega^{\rm QP}_{k})+\eta$,
where $\omega^{\rm QP}_k$ is the QP band energy at momentum $k$.  
Results at $k=(\pi/2, \pi/2)$ and $k=(0, 0)$ are shown in Figs.~\ref{fig:sigma_el} (a) and (b), respectively.  
Different $\alpha$ dependences are seen in the cases of $k=(\pi/2, \pi/2)$ and $k=(0, 0)$; 
$-\mathrm{Im} \Sigma(k; \omega^{\rm QP}_{k})$ decreases (increases) with increasing $\alpha$ at $k=(\pi/2, \pi/2)$ [$k=(0, 0)$], except for a vicinity of $\alpha=1$. 
This opposite behavior implies that the coherent QP motion at $k=(\pi/2, \pi/2)$ is promoted by quantum spin fluctuation, and the higher energy carriers at $k=(0, 0)$ are incoherently scattered by spin fluctuation. 

%%%%%%%%%%%%%%%%%%%%%%%%%%%%%%%%
%%%%%%%%%%%%%%%%%%%%%%%%%%%%%%%%
%\section{Transient dynamics}
%%%%%%%%%%%%%%%%%%%%%%%%%%%%%%%%

\subsection{One-particle excitation in transient states}

In this subsection, numerical results of the one-particle excitation spectra in the transient state are presented. 
As shown in  Sec~\ref{sec:gf}, the one-particle retarded Green's function is given in the perturbational expansion as 
$G^R(k; t, \tp)=\sum_{i=1}^4 G^R_i(k; t, \tp)$ where $G_1^R(k; t, \tp)=g^R(k; t, \tp)$, which is the retarded Green's function without dynamical doping, and $G^R_2(k; t, \tp)=0$. 
Then, we define difference of $G(k; \omega, \tp) $ from $g^R(k; \omega)$ as 
\begin{align}
\delta G^R(k; \omega, \tp)& \equiv G^R(k; \omega, \tp)-g^R(k; \omega) \nonumber \\
&=G^R_3(k; \omega,\tp)+G^R_4(k; \omega,\tp) , 
\end{align}
and a corresponding spectral weight as  
\begin{align}
\delta A(k; \omega, \tp)=- \frac{1}{\pi}\mathrm{Im}\delta G^R(k; \omega, \tp) , 
\end{align} 
where the sum rule 
\begin{align}
\int_{-\infty}^\infty \delta A(k; \omega,\tp) d \omega =0 , 
\label{eq:sumrule}
\end{align}
exists for any $\tp$. 
We have checked that Eq.~(\ref{eq:sumrule}) is satisfied in the numerical calculations within numerical errors. 

As for the momentum $k$ as a parameter of the one-particle Green's function and the momentum $\kz$ of the dynamically doped hole, we have some rules that 
$G_3=0$ when both $k$ and $\kz$ are on the $\rm X-M$ line, and that $G_4=0$ when either $k$ or $\kz$ is on this line. 
By taking into account these facts and the QP dispersion, 
we consider the following three cases in the numerical calculations: 
\par 
A)  $k=(0,0)$, $\kz=(0,0)$, \par 
B)  $k=(0,0)$, $\kz=(\pi/2,\pi/2)$, \par 
C)  $k=(\pi/2,\pi/2)$, $\kz=(0,0)$.

\begin{figure}[t]
\begin{center}
\includegraphics[width=\linewidth]{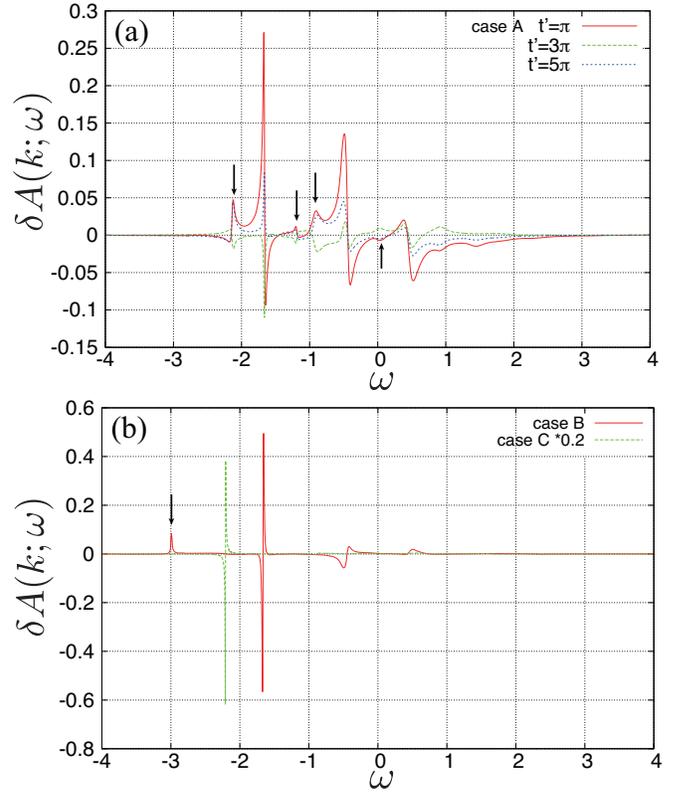}
\caption{
Time dependences of changes of the one-particle excitation spectra $\delta A(k;\omega)$. 
We chose $t'=\pi$,~$3\pi$, and $5\pi$. 
(a) results in case A) $[k=(0,0), \kz=(0,0)]$. 
Black arrows indicate ``magnon side peaks" (see the text). 
(b) results in case B) [$k=(0,0),~\kz=(\pi/2,\pi/2)$], and case C) with [$k=(\pi/2,\pi/2),~\kz=(0,0)$]. 
Results in case C) are multiplied by 0.2. 
Black arrow indicates ``exchange peak" (see the text). 
Other parameters are chosen to be $N =32$, $J=0.4$, $\alpha=0.8$, and $\eta=0.01$.
}
\label{fig:dGw}
\end{center}
\end{figure}

\begin{figure}[t]
\begin{center}
\includegraphics[width=\linewidth]{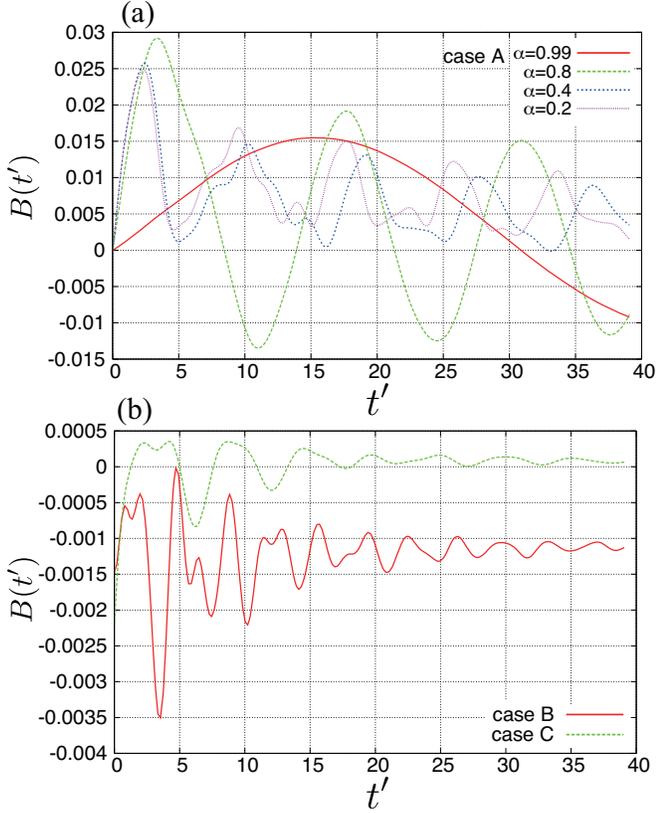}
\caption{
Time dependences of the integrated spectral weight $B(t)$ defined in Eq.~(\ref{eq:b}). 
(a) results in case A) $[k=(0,0), \kz=(0,0)]$, 
The anisotropy parameter is chosen to be $\alpha=0.99$, $0.8$, $0.4$, and $0.2$.   
(b) results in case B) [$k=(0,0),~\kz=(\pi/2,\pi/2)$], and case C) [$k=(\pi/2,\pi/2),~\kz=(0,0)$]. 
Other parameter values are chosen to be $N=32$, $J=0.4$, $\alpha=0.8$, and $\eta=0.01$.
}
\label{fig:B}
\end{center}
\end{figure}

Let us begin with case A). 
Figure~\ref{fig:dGw}(a) shows time dependences of $\delta A(\omega;\tp)$ in case A) $[k=(0,0)$, $\kz=(0,0)]$. 
Parameter values are chosen to be $J=0.4$ and $\alpha=0.8$. 
With increasing elapsed time $\tp$ after dynamical doping, 
remarkable changes appear around $\omega \simeq -1.6$, $-0.5$, and $0.4$,
which correspond to the lowest-three peak energies in ${\rm Im} g^R(k, \omega)$ before dynamical doping [see Fig.~\ref{fig:akw_el}(a)]. That is, the main three peaks shift to a lower energy side. 
In addition, small peaks, indicated by black arrows in Fig.~\ref{fig:dGw}(a), show oscillatory behaviors with time evolution. 
To examine origin of these transient changes in more detail,  
the calculated one-particle excitation spectra, as well as the optical conductivity spectra, are analyzed by using the dominant pole approximation (DPA) introduced in Ref.~\onlinecite{Kane}, where 
peaks in ${\rm Im} g^R(k; \omega)$ and ${\rm Im}d^R(k; \omega)$ are represented by a series of the Lorentz functions. 
Details are presented in Appendix \ref{sec:app:dpa}. 
As shown in Eq.~(\ref{eq:dpaG3}),
$\delta G^R(\omega,t^\prime)$ has poles at $\epsilon^i_k \pm \omega_{q=0}$, in addition to $\epsilon^i_k$ originated from ${\rm Im} g^R(k; \omega)$. 
These newly appearing peaks,  located around a sum of the original QP band energy and the magnon energy $\omega_{q}$, are termed ``magnon side peaks".

To examine the oscillatory behaviors shown above,  
we calculate the integrated spectral-weight change given by 
\begin{align}
B(k; t')=\int_{-\infty}^0 \delta A(k; \omega, t') d\omega. 
\label{eq:b}
\end{align}
Results in case A) for several values of the anisotropy parameter $\alpha$ are given in Fig.~\ref{fig:B}(a). 
Oscillations in $B(k; t')$ are clearly seen in the cases of large $\alpha$, 
and 
time period decreases with decreasing $\alpha$. 
From the analyses by DPA, 
the oscillations are characterized by $T_{sb} \equiv 2\pi/\omega_{q=0} \propto 1/(J\sqrt{1-\alpha^2})$ (see an exponential factor in Eq.~(\ref{eq:dpaG3})). 
In large $\alpha$, $B(k; t')$ has almost a single frequency component with $T_{sb}^{-1}$. 
This implies that the oscillating behavior originates almost only from the magnon side peaks indicated by arrows in Fig. \ref{fig:dGw}(a). 
On the other hand, 
multi-frequency oscillatory behavior in small $\alpha$ is attributed to the fact that 
other kind peaks have relevant contributions to $\delta A(k;t')$. 
One of the dominant contributions originate from peaks at $\omega=(\epsilon^b-\epsilon^c+\epsilon^d)-i(\Gamma^b+\Gamma^c+\Gamma^d)$ in Eq.~(\ref{eq:dpaG3}).  
This represents transitions between the inter ladder-type bands and are termed ``the exchange peaks". 

Next, we show the results in other cases of momenta $k$ and $\kz$. 
Results of $\delta A(k; \omega, t')$ and $B(k; t')$ are given in Fig.~\ref{fig:dGw}(b) and Fig.~\ref{fig:B}(b), respectively, 
in case B) [$k=(0,0)$, $\kz=(\pi/2,\pi/2)$] and case C) [ $k=(\pi/2,\pi/2)$, $\kz=(0,0)$]. 
As shown in Fig.~\ref{fig:dGw}(b), 
in both cases B) and C), dominant changes in $ \delta A(k, \omega)$ are seen as shifts of the main peaks in ${\rm Im}g^R(k, \omega)$, and contributions from the magnon side peaks are small. 
In case B), a small peak appears around $\omega \simeq -3$, indicated by a black arrow, 
which is below the lowest QP band before dynamical doping. 
Since a similar peak structure was observed in the previous calculations in the case with chemical doping~\cite{Plakida1994}, this is due to a carrier doping effect. 
As for the results of $B(k; t')$ in Fig.~\ref{fig:B}(b), clear oscillatory behaviors are not seen in both cases B) and C), unlike case A) with large $\alpha$. 
Damping of the oscillatory behavior is faster in case C) than that in case B). 
This is interpreted from the imaginary part of the self-energy before dynamically doping, that is, 
a damping rate $-{\rm Im} \Sigma(k;  \omega^{\rm QP}_{k})$ at $k=(0,0)$ is larger than that at $k=(\pi/2,\pi/2)$ for any values of $\alpha$ as shown in Fig.~\ref{fig:sigma_el}.

%%%%%%%%%%%%%%%%%%%%%%%%%%%%%%%%
\subsection{Optical conductivity in transient states}
\label{sec:OCinTS}

In this subsection, numerical results for the transient optical conductivity spectra after dynamical doping are presented. 
We note that no finite values in the optical conductivity spectra before dynamical hole doping in the $t-J$.
We focus on the $(xx)$ component of the optical conductivity spectra.

\begin{figure}
\begin{center}
\includegraphics[width=\linewidth]{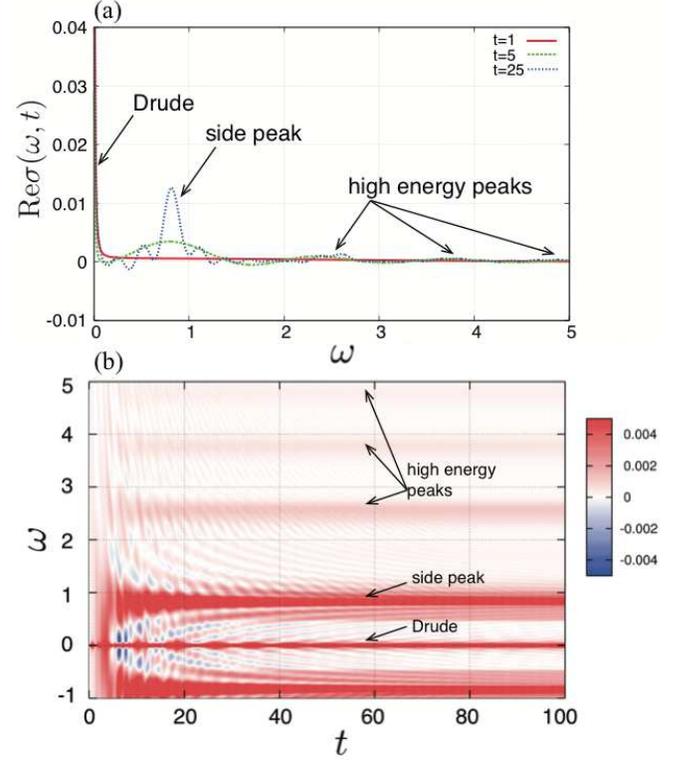}
\caption{
Transient optical conductivity spectra at several times; 
$t=1$ (red), $5$ (green), and $25$ (blue). 
A momentum of the dynamically doped hole is chosen to be $\kz=(\pi/2, \pi/2)$.
(b) a contour plot of the optical conductivity spectra in a plane of time and frequency. 
Other parameters are chosen to be $J=0.4$ and $\alpha=0.8$.
}
\label{fig:sigma1}
\end{center}
\end{figure}

First, we present results where the momentum of the dynamically doped hole is chosen to be $\kz=(\pi/2,\pi/2)$, which corresponds to the lowest QP peak in ${\rm Im} g^R(k; \omega)$. 
In Fig.~\ref{fig:sigma1}(a), the time dependences of the real parts of the optical conductivity spectra are presented. 
Just after the dynamical hole doping at $t=0$, only the Drude peak is confirmed at $t=1$. 
After a lapse of time, sub peak structures appear and grow around $\omega=0.8$, $2.5$ and other energies. 
A peak at around $0.8$ is termed the ``side peak", and 
other peaks located around $2.5$, $3.8$ and others are termed ``high-energy peaks", from now on. 
As explained later, emergence of these peaks implies that dynamically doped hole is dressed by magnons. 
From the analyses by DPA, 
the Drude and finite frequency peaks grow as $t^2$ and $t^3$, respectively, 
as shown in Eq.~(\ref{eq:t2t3}).  

A contour plot of $\sigma(\omega, t)$ in a $\omega-t$ plane is presented in Fig.~\ref{fig:sigma1}(b).
Time evolutions of the side peak and the high energy peaks are clearly shown.   
Origins of the side peak and the high energy peaks are clarified by DPA. 
In Eq.~(\ref{eq:app:hq}), 
these peaks are characterized by poles of $h_{k_0,q}$, introduced in Eq.~(\ref{eq:sigma_h}), at $\omega=\pm(\epsilon_{\kz-q}^b-\epsilon_\kz^c+\omega_q)+i(\Gamma_{\kz-q}^b-\Gamma_\kz^c+\eta)$, 
where $\epsilon_k^i$ and $\Gamma_k^i$ are energy and a damping factor for the $i$-th peak of ${\rm Im} g^R(k; \omega) $.
The side peak and the high-energy peaks originates from poles with $b=c$ and $b\ne c$, respectively, and are attributed to the intra-band transitions inside the lowest-energy QP band, and the inter-band transitions between the ladder-like multiple-bands, respectively [see Fig.~\ref{fig:akw_el}(a)]. 
Thus, energies of the side peak and the high energy peaks are mainly dominated by band widths of the QP band 
and energy differences between the ladder-like peaks, respectively. 
To clarify roles of magnon on the optical conductivity in more detail, we introduce a hypothetic situation that 
the self-energy for the hole Green's function given in Eq.~(\ref{eq:scba}) is set to be zero artificially. 
In this case,  the QP band in ${\rm Im}g^R$ does not show a dispersion, and magnon only appears in the current vertex. We observe that the high energy peaks do not appear in the optical conductivity,  
and the energy of the side peak depends only on the magnon energy and amplitude of the current vertex.  
From these results, the time evolution of the optical conductivity is interpreted that a dynamically doped hole is dressed by magnon cloud with time evolution.

\begin{figure}
\begin{center}
\includegraphics[width=\linewidth]{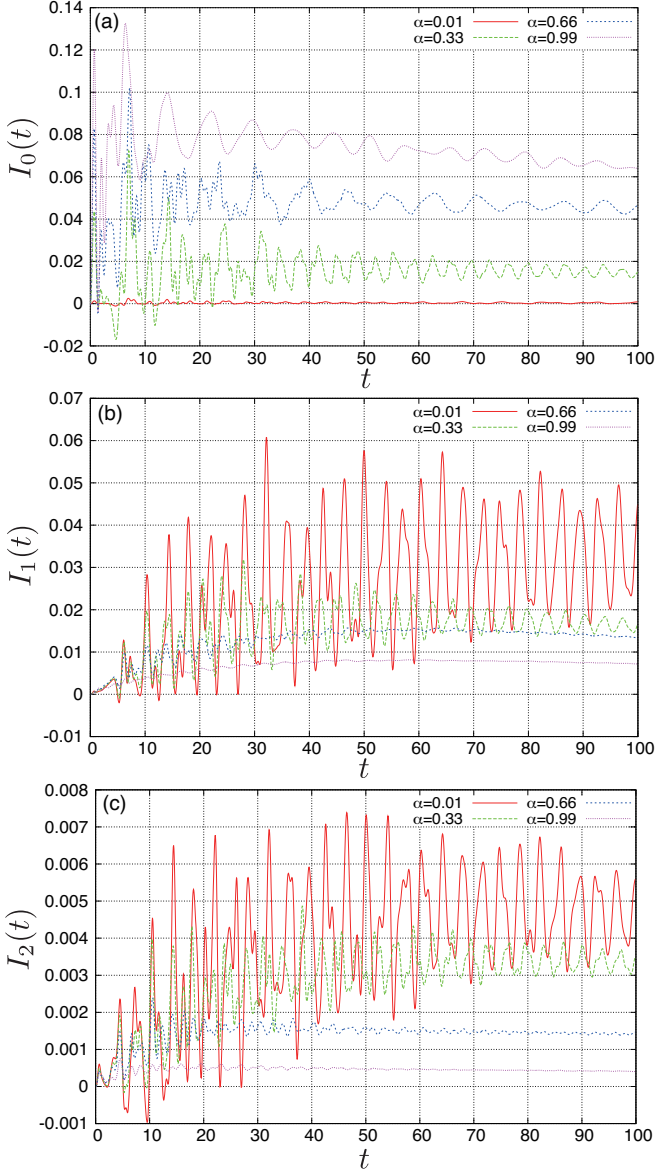}
\caption{
Time dependences of the peak intensities $I_i(t)$ defined in Eq.~(\ref{eq:integ_peak}) 
for several values of the anisotropy parameter $\alpha$. 
(a) the Drude peak $(i=0)$, (b) the side peak $(i=1)$, and (c) the high-energy peak ($i=2$). 
Parameters are chosen to be  $\kz=(\pi/2,\pi/2)$ and $J=0.4$. 
Values for $\omega_i$ and $\delta_i$ in Eq.~(\ref{eq:integ_peak}) are taken to be 
$(\omega_0, \delta_0, \omega_1, \delta_1, \omega_2, \delta_2)=$
$(0, 0.01, 0.78, 0.03, 2.57, 0.02)$ for $\alpha=0.01$, 
$(0, 0.01, 0.81, 0.07, 2.60, 0.05)$ for $\alpha=0.33$, 
$(0, 0.01, 0.83, 0.07, 2.58, 0.12)$ for $\alpha=0.66$, and 
$(0, 0.01, 0.84, 0.05, 2.56, 0.04)$ for $\alpha=0.99$. 
}
\label{fig:sigma3}
\end{center}
\end{figure}

Detailed time evolution for each peak is examined by introducing the integrated peak intensity  around a peak position. 
This is defined by 
\begin{align}
I_i(t)
=
\int_{\omega_{i}-\delta_i}^{\omega_{i}+\delta_i
}
d\omega
\mathrm{Re}\sigma(\omega,t) , 
\label{eq:integ_peak}
\end{align}
where peak is identified by an index $i$, 
$\omega_i$ is a peak position at $t=100$, 
and $\delta_i$ is width in integrated energy region. 
Numerical values of $\delta_i$ are chosen to be an energy where peak intensities are almost zero.
We define that $i=0$ and 1 are for the Drude peak, and the side peak, respectively, 
and $i\ge 2$ are for the high energy peaks. 
Time dependences of $I_i(t)$ are shown in Figs.~\ref{fig:sigma3}(a), (b) and (c) for 
$i=0$ (Drude peak), $i=1$ (side peak) and $i=2$ (high energy peak), respectively. 
Strong oscillatory behaviors are seen in all peaks, in particular, fast remarkable oscillations in small $\alpha$. 
These are attributed to ladder-like multiple peaks in ${\rm Im}g^R(k, \omega)$, 
and are identified as the ``coherent magnon oscillations".

Deferent $\alpha$ dependences of the intensity are seen in the Drude and other peaks; 
$I_0(t)$ [$I_1(t)$ and  $I_2(t)$] decreases (increase) with increasing  the anisotropic parameter $\alpha$. 
These different $\alpha$ dependences are interpreted as follows. 
As shown in Eqs.~(\ref{eq:red})-(\ref{eq:sigma_h}) and Eq.~(\ref{eq:app:hq}), 
intensities of the side peak and the high energy peaks are given by the imaginary parts of $h_{{\ki}, q}(t,t^\prime)$. Since poles are located at $\omega=\pm(\epsilon_{bc}+\omega_q)$ which give a relaiton between the magnon and hole monenta, these intensities are sensitive to the band dispersions of magnon and hole. 
On the other side, the Drude peak is given by the real part of $h_{{\ki}, q}(t,t^\prime)$, which is related to integrated ${\rm } h_{{\ki}, q}(t,t^\prime)$ in terms of energy. 
Therefore, the Drude peak intensity is not sensitive to details of the dispersions of hole and magnon, 
but to the coupling constants $M_{kq}$ [Eq.~(\ref{eq:mkq})] and ${\bm V}_{kq}$ [Eq.~(\ref{eq:vkq})], which directly depend on $\alpha$. 

\begin{figure}
\begin{center}
\includegraphics[width=\linewidth]{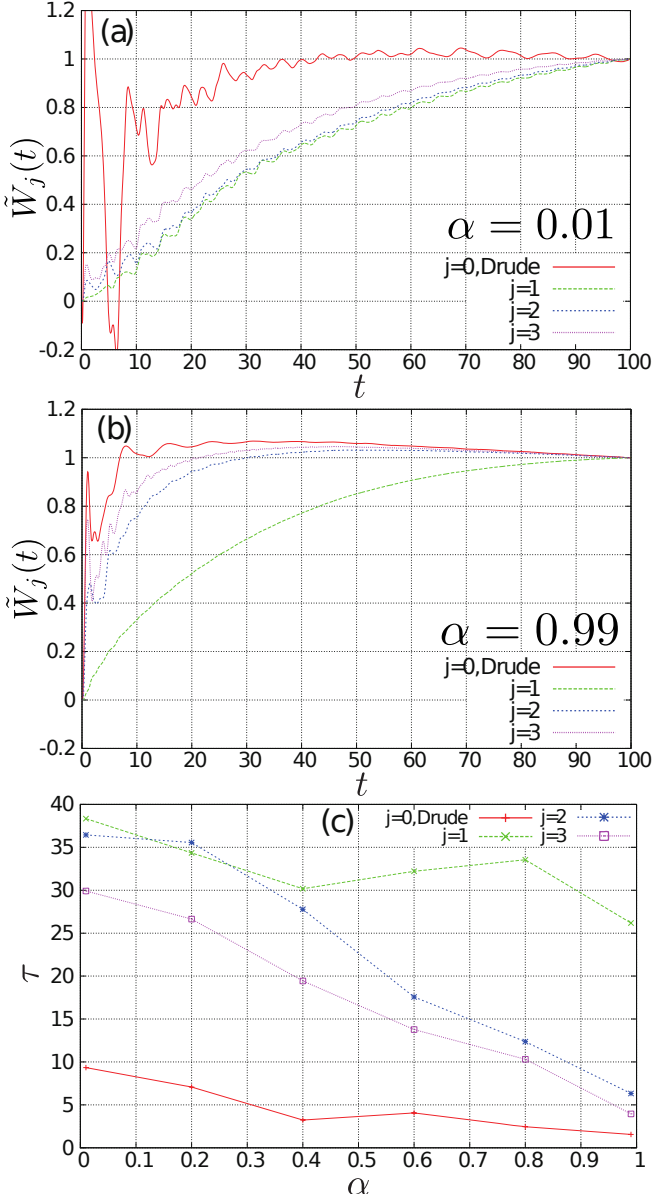}
\caption{
Time dependences of $W_i(t)$ defined in Eq. (\ref{eq:relax}) normalized at $t=100$, i.e. 
$\widetilde{W}_i(t)=W_i(t)/W_i(100)$, for each peak. 
The anisotropy parameter is chosen to be $\alpha=0.01$ in (a), and $0.99$ in (b).  
(c) relaxation time as functions of $\alpha$ for each peak. 
Parameter value is chosen to be $J=0.4$. 
}
\label{fig:sigma5}
\end{center}
\end{figure}
Next, slow dynamics, without fast oscillatory behavior, are examined by integrating $I_i(t)$ with respect to time since dynamical doping as 
\begin{align}
W_i(t)
=
\frac{1}{t}
\int_0^t
dt^\prime
I_i(t^\prime) . 
\label{eq:relax}
\end{align}
In Fig~\ref{fig:sigma5}, 
results normalized by $W_i(t)$ at $t=100$, i.e. 
$\widetilde{W}_i(t)=W_i(t)/W_i(100)$, are shown. 
The anisotropy parameter is chosen to be $\alpha=0.01$ and $0.99$ in Fig.~\ref{fig:sigma5} (a) and (b), respectively. 
Each data are smoothly converged to 1 in large $t$. 
That is, the dynamically doped hole gradually undergoes change into a quasi-steady state with coherent oscillations. 
Relaxation times depend on peaks and $\alpha$; 
fast relaxations are seen in the Drude peak $(i=0)$ with $\alpha=0.01$, 
and the Drude and side peaks ($i=0$ and $1$) with $\alpha=0.99$. 
We fit the time dependences of ${\widetilde W}_i(t)$ by a simple exponential function $1-\exp(-t/\tau)$, 
and show relaxation times in Fig.~\ref{fig:sigma5}(c). 
The $\alpha$ dependence is remarkable in the high energy peak ($i=2$ and $3)$; 
the relaxation is slow (fast) in the Ising (Heisenberg)-like case. 
As explained previously, the high energy peak is attributed to the 
inter-band transition between the ladder-like multiple-bands in ${\rm Im} g^R(q, \omega)$. 
It is interpreted that the scattering probability is large (small) in the small (large) energy difference between the bands for large (small) $\alpha$. 
Remarkable incoherent backgrounds in ${\rm Im} g^R$ in large $\alpha$ induce additional transition pathways.

%%%%%%%%%%%%%%%%%%%%%%%%%%%%%%%%
%\subsection{Optical conductivity : initial states' $\ki$ dependence}
\begin{figure}
\begin{center}
\includegraphics[width=\linewidth]{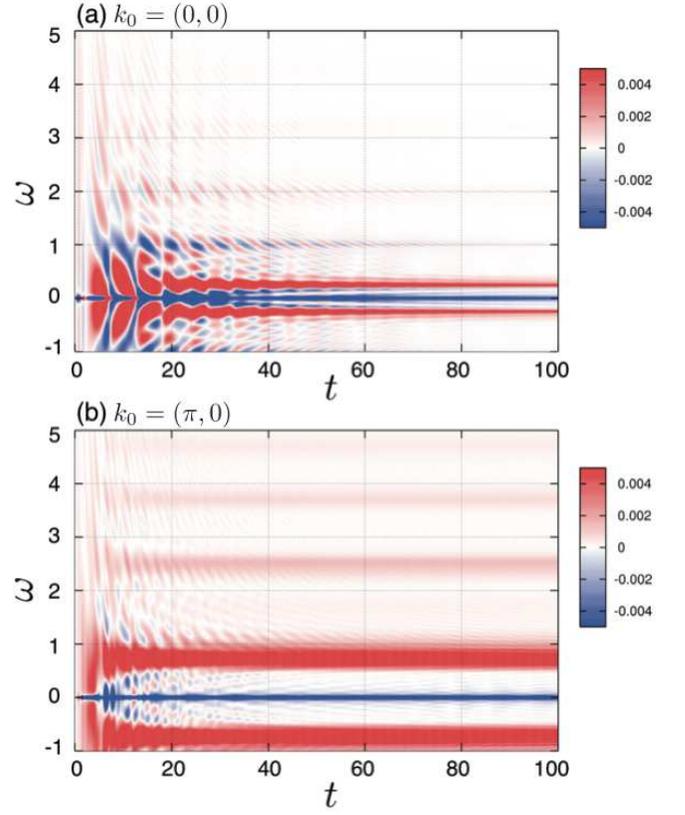}
\caption{
Contour plots of the optical conductivity spectra in the cases of 
$\kz=(0,0)$ in (a), and $\kz=(\pi,0)$ in (b). 
Parameter values are chosen to be $J=0.4$, and $\alpha=0.8$. 
}
\label{fig:sigma6}
\end{center}
\end{figure}

\begin{figure}
\begin{center}
\includegraphics[width=\linewidth]{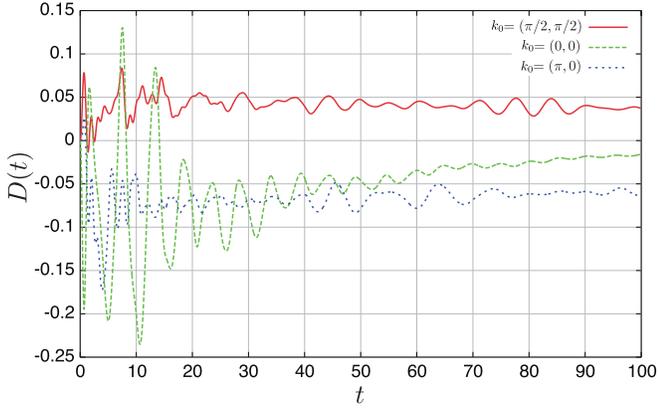}
\caption{
Time dependence of the Drude weight $D(t)$ for several momenta of the dynamically doped holes. 
Parameter values are chosen to be $J=0.4$ and $\alpha=0.8$. 
Bold, broken and dotted lines are for $k_0=(\pi/2, \pi/2)$, $(0, 0)$ and $(\pi, 0)$, respectively. 
}
\label{fig:sigma7}
\end{center}
\end{figure}

Finally, we show results for other momenta of dynamically doped hole. 
In Fig.~\ref{fig:sigma6} (a) and (b), we present the optical conductivity spectra $\mathrm{Re}\sigma(\omega,t)$ for $\kz=(0,0)$ and $(\pi,0)$, respectively. 
In both cases, the Drude peak, the side peak and the high energy peaks are confirmed. 
The energy of the side peak for $\kz=(0, 0)$ is lower than those for $\kz=(\pi/2,\pi/2)$ and $(\pi,0)$.
This is explained by DPA that 
the energy of the side peak is dominated by a pole at $\omega=\epsilon^b_{\kz-q}-\epsilon^c_{\kz}+\omega_q$ with $b=c$, where $\epsilon^b_{\kz-q}-\epsilon^c_{\kz}$ is negative for any $q$ in the case of $\kz=(0,0)$.
This $k_0$ dependence of the side peak energy is remarkable in large $\alpha$ where the dispersions of magnons and holes are large.

It is worth to note that 
Figure~\ref{fig:sigma6}~(b) looks similar to Fig.~\ref{fig:sigma1}(b), for the side peak and the high energy peak. 
This is because the QP band in ${\rm Im}g^R (k, \omega)$ is nearly degenerated along  $k=(\pi/2,\pi/2)$ and $(\pi,0)$.
The most remarkable deference between Fig.~\ref{fig:sigma6}(b) and Fig.~\ref{fig:sigma1}(b) is seen in a sign of the Drude weight, which is negative in Fig.~\ref{fig:sigma6}(b). 
More detailed time profiles of $D$ for several $\kz$ are presented in Fig.~\ref{fig:sigma7}.
Negative values for $k_0=(\pi,0)$ and $(0,0)$ tend to be positive gradually.
This negative value is interpreted from a population inversion and light emission. 
As seen in Fig~\ref{fig:akw_el}(a), the second derivative of the QP band dispersion with respect to the momentum, $\partial^2\omega_k^{\mathrm{QP}}/\partial k^2$, 
is negative at $k=(\pi,0)$ and $(0,0)$ and positive at $k=(\pi/2,\pi/2)$.
This convexity of the QP band in ${\rm Im} g^R(k; \omega)$ for a dynamically doped hole determines a sign of $D$.

%%%%%%%%%%%%%%%%%%%%%%%%%%%%%%%%
%%%%%%%%%%%%%%%%%%%%%%%%%%%%%%%%
\section{Conclusion}

In conclusion, we have studied real-time hole dynamics injected into a Mott insulator with antiferromagnetic long range order. The transient optical conductivity spectra and one-particle excitation spectra are formulated 
by applying the Keldysh Green's function formalism and the self-consistent Born approximation to the two-dimensional $t-J$ model.  
Here we list the main results. 
At early stage just after dynamical hole doping, the Drude component only appears, and then the finite-energy incoherent components, which are termed the side peak and the higher-energy peaks gradually grow [see Figs.~\ref{fig:sigma1}(a) and (b)]. 
The side peak and higher energy peaks are identified as the intra- and inter-band transitions of the ladder-type bands, respectively, associated with magnon excitations. 
In a small $t$ region, intensities of the Drude and the incoherent parts grow as $t^2$ and $t^3$, respectively. 
These time evolutions imply that a dynamically doped bare hole is gradually dressed by magnon cloud, and 
undergo changes into spin polaron QP state. 
Time profile of the Drude and incoherent peaks show fast oscillatory components, i.e. coherent magnon oscillation, and slow relaxation dynamics scaled by an exponential function (see Fig.\ref{fig:sigma3}). 
The fast component is remarkable in the small $\alpha(\equiv J_{\perp}/J_{\parallel})$ case (Ising-like case), where magnon dispersion is almost flat and provides a single time scale. 
As for the slow relaxation dynamics, only the higher energy peaks show strong $\alpha$ dependence 
[see Fig.\ref{fig:sigma5}(c)]. 
This is interpreted from the fact that the inter-band hole transitions by magnon are effective in the large $\alpha$ case (Heisenberg-like case) where magnon band width is large. 

The present numerical calculations simulate early stage in the time-resolved pump-probe experiments. 
When we take $t=0.5$eV, time scale $t=$30 in the present calculation corresponds to about 40 femtosecond, which is a realistic time scale in recent ultrafast optical experiments. 
Our calculations predict a time lag for observation of the incoherent component of the optical conductivity spectra 
associated with a coherent magnon oscillation. 
Purely electronic processes of dynamically doped carriers are detected in careful measurements and analyses of early stage of time-resolved optical spectra, where dynamics of electron and hole carriers are able to be separated.~\cite{Okamoto1} 
More direct test of the present simulations will be performed by the photo-carrier injection into the heterostructure where holes or electrons are selectively introduced into a sample.~\cite{Hiroi,Muraoka,Yada1,Yada2} 

It is worth to mention that the present formalism is also applicable to real-time carrier dynamics doped into Mott insulators with the orbital degree of freedom. 
There is a number of studies of photo-irradiation effects in orbitally degenerate correlated electron systems.~\cite{Polli,Mazurenko,Beaud,Ichikawa,Maeshima}
A doped Mott insulator with orbital degree of freedom is described by the $t-J$ type model,~\cite{Mack,Ishihara_scb} where 
a similar decomposition of the restricted hole operator [see Eqs.~(\ref{eq:Ht}) and (\ref{eq:Hj})] is justified;  
doped carrier dynamics are represented by a spin- and orbital-less fermion which interacts with magnon and ``orbiton". 
One of the qualitative difference of an orbitally degenerate Mott insulator from a present single band Mott insulator in this scheme is the fact that an occupied electron number in a specific orbital is not conserved.~\cite{Ishihara_scb} 
This fact brings about a free kinetic term for fermion as 
${\cal H}_t \sim \sum_{k} h_k^\dagger h_k$ where orbiton operators does not appear, in contrast to Eq.~(\ref{eq:Ht}). Expected qualitatively different real-time  carrier dynamics will be examined in near future. 

\begin{acknowledgments}
Authors thanks H. Okamoto, S. Koshihara and S. Iwai for helpful discussions. 
This work was supported by ``Core Research Evolutional and Science and Technology" by JST,  
KAKENHI from MEXT and Tohoku University ``Evolution'' program. 
Parts of the numerical calculations are performed in the supercomputing systems in ISSP,  the University of Tokyo. 
\end{acknowledgments}

\appendix
%%%%%%%%%%%%%%%%%%%%%%%%%%%%%%%%
%%%%%%%%%%%%%%%%%%%%%%%%%%%%%%%%
\section{Dominant pole approximation}
\label{sec:app:dpa}
In this Appendix, we present formalisms and results of the dominant pole approximation introduced in Ref.~\onlinecite{Kane} and utilized in Sec.~\ref{sec:results}.
As shown in the results in Refs.~\onlinecite{Kane,Martinez} and Fig.~\ref{fig:akw_el}, 
the calculated one-particle excitation spectra consist of a series of multiple poles, and incoherent background. 
We assume in this approximation that the formers are represented by Lorentzian peaks, and a later is neglected.
The hole-Green's function in vacuum is given as 
\begin{align}
\label{eq:dpa:g}
g^R(k; t-t^\prime)
&=
-i \sum_a z_{k}^a e^{\sbra{-i\epsilon_{k}^a-\Gamma_{k}^a}\sbra{t-t^\prime}}\theta(t-t^\prime),
\end{align}
where a superscript $a$ identifies a peak, and 
$z_{k}^a$, $\epsilon_{k}^a$, and $\Gamma_{k}^a$ are weight, energy, and a damping factor of the peak $a$ with momentum $k$, respectively.
In the same manner, the magnon Green's function defined in vacuum is given as  
\begin{align}
d^R(q;t-t^\prime)
&=
-i e^{\sbra{-i\omega_q-\eta}\sbra{t-\tp}}\theta(t-\tp) ,
\end{align}
where $\omega_q$ and $\eta$ are energy and a damping factor of magnon, respectively. 

%%%%%%%%%%%%%%%%%%%%%%%%%%%%%%%%
%\subsection{Green's function}
%\label{sec:app:dpa:g}

In this approximation, 
$G_3^R(k; \omega,\tp)$ introduced in Eq.~(\ref{eq:g3}) is explicitly obtained as 
\begin{align}
G_3^R(k; \omega,\tp)
=&
\sum_{a,b,c,d} \sum_{\sigma=\pm1}
\frac{\sigma C_{abcd}}{-\epsilon_{\kz}^c+\epsilon_{\kz}^d-\sigma\omega_{q=0}-i\sbra{\Gamma_{\kz}^c+\Gamma_{\kz}^d-\eta}} 
\nonumber \\
&\times
\frac{1}{\omega-\epsilon_{k}^a+i\Gamma_{k}^a}
\nonumber \\
&\times
\Biggl \{
\frac{{\rm exp} \left[i\sbra{\epsilon_{\kz}^c-\epsilon_{\kz}^d}-\sbra{\Gamma_{\kz}^c+\Gamma_{\kz}^d}  \tp \right]}
{\omega-\sbra{\epsilon_{k}^b-\epsilon_{\kz}^c+\epsilon_{\kz}^d}+i\sbra{\Gamma_{k}^b+\Gamma_{\kz}^c+\Gamma_{\kz}^d}}
\nonumber \\
&\ \ 
-\frac{e^{\sbra{-i\sigma\omega_{q=0}-\eta}\tp}}{\omega-\sbra{\epsilon_{k}^b+\omega_0\sigma}+i\sbra{\Gamma_{k}^b+\eta}}
\Biggr \}  \label{eq:dpaG3} \\
=&
\sum_{a,b,c,d} \sum_{\sigma=\pm 1}
\frac{\sigma C_{abcd}e^{\sbra{-i\sigma\omega_{q=0}-\eta}\tp}}{\omega-\epsilon_k^a+i\Gamma_k^a} \nonumber \\
& \times 
\frac{D^\sigma_b\sbra{\omega,\Omega^\sigma_{cd},\tp}-D^\sigma_b\sbra{\omega,0,\tp}}{\Omega^\sigma_{cd}} , 
\end{align}
where 
\begin{align}
D^\sigma_b\sbra{\omega,z,\tp}
=&
\frac{e^{iz\tp}}{\omega-\sbra{\epsilon_{k}^b+\sigma\omega_{q=0}}+i\sbra{\Gamma_{k}^b+\eta}-z} , 
\end{align}
and 
\begin{align}
\Omega^\sigma_{cd}=-\epsilon_{\kz}^c+\epsilon_{\kz}^d-\omega_0\sigma-i\sbra{\Gamma_{\ki}^c+\Gamma_{\kz}^d-\eta} . 
\end{align}
We also obtain $G_4^R(k; \omega, \tp)$ introduced in Eq.~(\ref{eq:g4}) as 
\begin{align}
G_4^R(k; \omega,\tp)
=&
\sum_{a,b,c,d} \sum_{\sigma=\pm1}
\frac{-\sigma\widetilde{C}_{abcd}^\sigma}{\omega-\epsilon_{k}^a+i\Gamma_{k}^a}
\nonumber \\
&\times 
\frac{e^{\left \{i\sbra{\epsilon_{\kz}^c-\epsilon_{\kz}^d}-\sbra{\Gamma_{\kz}^c+\Gamma_{\kz}^d}\right \}\tp}}{\omega-\sbra{\epsilon_{k}^b-\epsilon_{\kz}^c+\epsilon_{\kz}^d}+i\sbra{\Gamma_{k}^b+\Gamma_{\kz}^c+\Gamma_{\kz}^d}}
\nonumber \\
& \times
\frac{1}{\omega-\sbra{\epsilon_{\kz}^d+\sigma\omega_{k-\kz}}+i\sbra{\Gamma_{\kz}^d+\eta}} . 
\label{eq:dpaG4}
\end{align}
Poles in ${\rm Im} g^R(k; \omega)$ are labeled by indices $a$, $b$, $c$ and $d$. 
We introduce constants 
\begin{align}
C_{abcd}
=&
\frac{M_{k0}M_{\kz 0}}{2}z^a_k z^b_k z^c_{\kz} z^d_{\kz} , 
\end{align}
and 
\begin{align}
\widetilde{C}_{abcd}^\sigma
=&
\frac{M_\sigma^2}{2}z^a_k z^b_k z^c_\kz z^d_{\kz} , 
\end{align}
with 
\begin{align}
M_{\sigma=1}
=&
M_{k,k-\kz},
\\
M_{\sigma=-1}
=&
M_{\kz,\kz-k}.
\end{align}
An index $\sigma$ represents a direction of the magnon propagator, i.e. 
 $\sigma=1$ for $d(q,t)$ and $\sigma=-1$ for $d^*(q,t)$, and we have 
\begin{align}
d(0;t)+d^*(0;t)
=&
\sum_{\sigma=\pm 1} (-i)\sigma e^{\sbra{-i\omega_0\sigma-\eta}t} . 
\end{align}
We note that $G_3^R$ and $G_4^R$ cancel with each other, when $t^\prime=0$ and $k=\kz$.

Poles in $G_3(k; \omega, \tp)$ and $G_4(k; \omega, \tp)$ are classified as the following three types: 
\par \noindent
i) $\omega=\epsilon^a-i\Gamma^a$, originating from the poles in $g^R(k; \omega)$. 
\par \noindent
ii) $\omega=(\epsilon^b+\omega_q\sigma)$, representing ``magnon side peaks". 
This peak oscillates by a factor $e^{(-i\sigma\omega_{q=0}-\eta)t^\prime}$
and is characterized by a period $T_{\mathrm{sb}}=2\pi/\omega_{q=0}$.
\par \noindent 
iii) $\omega=\sbra{\epsilon^b-\epsilon^c+\epsilon^d}-i(\Gamma^b+\Gamma^c+\Gamma^d)$,
termed ``exchange peaks". 
Time evoluton of this peak is dominated by a factor 
${\rm exp} \left [ -i \left \{( \epsilon_\kz^c-\epsilon^d_\kz)-(\Gamma_\kz^c+\Gamma_\kz^d) \right \} t^\prime \right ]$.
Decay and oscillations are governed by $\tau_{\mathrm{decay}}=1/(\Gamma_c+\Gamma_d)$
and $T_{\mathrm{ex}}=2\pi/(\epsilon_c-\epsilon_d)$, respectively, 
when the transition occurs between different bands ($c\neq d$).

As shown in Fig.~\ref{fig:sigma_el}, with decreasing $J$, the damping rate $\tau_{\mathrm{decay}}$ decreases. 
When $\tau_{\mathrm{decay}}$ is smaller than the period $T_{\mathrm{ex}}$, the exchange peaks decay rapidly.
On the other side, in the case of large $J$ where the ladder-type peaks are well separated,  
$\tau_{\mathrm{decay}}$ is large, and contribution from the ``exchange peaks" is not negligible.

%%%%%%%%%%%%%%%%%%%%%%%%%%%%%%%%
%\subsection{Optical conductivity}
%\label{sec:app:dpa:sigma}

The optical conductivity spectra are also 
evaluated in DPA. 
A part of the optical conductivity spectra, $h_{\kz,q}(t,t^\prime)$ defined in Eq.(\ref{eq:sigma_h}),
is given as 
\begin{align}
h_{\kz,q}(t,t^\prime)
&=
-\sum_{a,b,c} z_{\kz}^a z_{\bar{l}}^b z_{\kz}^c \sum_{\sigma=\pm 1}\frac{\sigma}{2 i}
\nonumber \\
&\times
{\rm exp} \left[\sbra{i\sigma\epsilon_{ac}-\Gamma_{ac}}t \right] 
\nonumber \\ &\times
{\rm exp} \left[\sbra{i\sigma\epsilon_{bc}+i\omega_q-\gamma_{bc}-\eta}(t-t^\prime) \right],
\end{align}
where $\bar{l}=\kz-q$, $\epsilon_{ac}=\epsilon_{\kz}^a-\epsilon_{\kz}^c$, $\epsilon_{bc}=\epsilon_{\kz-q}^b-\epsilon_{\kz}^c$, $\Gamma_{ac}=\Gamma_{\kz}^a+\Gamma_{\kz}^c$, and $\gamma_{bc}=\Gamma_{\bar{l}}^b-\Gamma_{\kz}^c$. 
A Fourier transform of $h_q(t,t^\prime)$ in terms of $t-t'$ is obtained as 
\begin{align}
h_{\kz,q}(\omega,t)
&=
\sum_{abc}
\frac{z_{\kz}^az_{\bar{l}}^b z_{\kz}^c}{2}
\sum_{\sigma=\pm 1}
\sigma
e^{(i\sigma\epsilon_{ac}-\Gamma_{ac})t}
\nonumber \\
& \times
\frac{
{\rm exp} \left [ \sbra{i\omega-i\sigma(\epsilon_{bc}+\omega_q)-(\gamma_{bc}+\eta)}t \right ]
-1
}{
\omega-\sigma(\epsilon_{bc}+\omega_q)+i(\gamma_{bc}+\eta)
}, 
\label{eq:app:hq}
\end{align}
where poles exist at $\omega=\pm(\epsilon_{bc}+\omega_q)-i(\gamma_{bc}+\eta)$.
The width of the peak at finite frequencies is determined by the damping factor $\gamma_{bc}$ and dispersion of the excitation energy $\epsilon_{bc}+\omega_{q}$.
Among several terms with respect to the indexes $abc$, 
contributions where $a$ and $c$ belong to the lowest QP band in the one-particle excitation spectra are large, because of the damping factors $e^{-\Gamma_{ac}t}$ which makes other terms small.
We classify the peaks at finite frequency in the optical conductivity spectra by 
\par \noindent
i) $b=c$ in which the peak is termed the ``the side peak".
\par \noindent
ii) $b\neq c$ in which the peak is termed ``the high energy peaks". 
\par
In the case i), scatterings of the dynamically doped hole occur from $\kz$ to $\kz-q$ inside of the lowest QP band, and induce a peak in the optical conductivity spectra at around $\omega=\pm(\delta \epsilon+\omega_q)$, where $\delta\epsilon$ is characterized by the band width of hole.
As a dynamically doped hole is dressed by spin clouds, 
$\delta\epsilon$ as well as an energy of the side peak increase. 
In the case ii), 
transitions of the dynamically doped hole from the lowest energy QP band to the higher energy bands occur. 

From Eq.~(\ref{eq:app:hq}), early time dynamics in the optical conductivity spectra are estimated. 
By expanding the right hand side in Eq.~(\ref{eq:app:hq}) with respect to $t$, we have 
\begin{align}
h_{\kz,q}(\omega,t)
&=
\sum_{abc}\frac{ z_{\kz_a} z_{\bar{l}_b} z_{\kz_c}}{2}\epsilon_{ac}t^2+O(t^3),
\label{eq:t2t3}
\end{align}
where the coefficient of $O(t^3)$ is a complex number and finite for any $\omega$.
The Drude weight, being proportional to $\mathrm{Re}[h_q(\omega=0,t)]$, 
obeys $t^2$, and 
finite frequency component $\sigma(\omega\neq 0,t)$, being proportional to $\mathrm{Im}[h_q(\omega\neq 0,t)]$, 
obeys $\propto t^3$.

\noindent
$\ast$Present address: Department of Basic Science, The University of Tokyo, 3-8-1 Komaba, Meguro-ku, Tokyo, 153-8902 Japan.

\end{document}